\documentclass{elsart}
\usepackage{graphicx}
\begin{document}
\begin{frontmatter}

\title{System size stochastic resonance in a model for opinion formation}
\author{Claudio J. Tessone}\ead{\tt tessonec@imedea.uib.es},
\author{Ra\'ul Toral}\ead{\tt raul@imedea.uib.es}

\address{Institut Mediterrani d'Estudis Avan\c{c}ats IMEDEA
(CSIC-UIB), Ed. Mateu Orfila, Campus UIB, E-07122 Palma de Mallorca, Spain}

\date{\today}
\begin{abstract}We study a model for opinion formation which incorporates three basic ingredients for the evolution of the opinion held by an individual: imitation, influence of fashion and randomness. We show that in the absence of fashion, the model behaves as a bistable system with random jumps between the two stable states with a distribution of times following Kramer's law. We also demonstrate the existence of system size stochastic resonance, by which there is an optimal value for the number of individuals $N$ for which the average opinion follows better the fashion. 
\end{abstract}
\end{frontmatter}

\section{Introduction}

It is nowadays well established that the stochastic terms (noise) in the equations of motion of a dynamical system can have a constructive effect leading to some sort of order. An example is the appearance of ordered phases in a scalar field theory when the noise intensity is increased~\cite{BPT94,IGTS01,OS99}. The classical prototype is that of {\sl stochastic resonance}~\cite{JSP70,GHJM98,BSV81,NN81} by which an adequate value of the noise intensity helps to synchronize the output of a nonlinear dynamical system with an external forcing. Amongst other examples, that of {\sl coherence resonance} shows again that the proper amount of noise helps to improve the regularity of the output of an excitable~\cite{PK97} or chaotic~\cite{PTMCG01} system. Similar results have been referred to as {\sl stochastic coherence} in~\cite{ZGBU03} or {\sl stochastic resonance without external periodic force}~\cite{GDNH93,RS94}. In these examples, the subtle interaction between the nonlinear terms, the coupling (either internal or with an external source) and the noise produce the desired effect. Most of these previous works have considered the appearance of order as a result of tuning the noise intensity to its proper value, whereas the role of system size has been either neglected, or analyzed in terms of standard finite-size theory for phase transitions~\cite{BPTK97}.

A recent line of work, however, considers that the output of a nonlinear stochastic system can have a nontrivial dependence on its size (or number of constituents). Some recent work on biological models~\cite{SGH01,JS01,SJ02} consider Hodgkin-Huxley type models to show that the ion concentration along biological cell membranes displays (intrinsic) stochastic resonance as well as coherence resonance as the number of ion channels is varied. These references also discuss the possible biological implications. A similar result~\cite{TMG03} shows that in the absence of external forcing, the regularity of the collective output of a set of coupled excitable FitzHugh-Nagumo systems is optimal for a given value of the number of elements. This is a {\sl system size stochastic coherence} effect. 

In physical systems, {\sl system size resonance}~\cite{PZC02} has been found in the Ising model, as well as in a set of globally (or local) coupled generic ($\phi^4$-type) bistable systems $(x_1,\dots,x_N)$ under the influence of an external periodic forcing and uncorrelated Gaussian white-noises $\xi_i(t)$:

\begin{eqnarray}\label{eqxi}
\dot{x_{i}} &=&x_{i}-x_{i}^3+\frac{K}{N}\sum_{j=1}^N 
(x_{j}-x_{i})  + \sqrt{D} \xi_i(t) + A \cos (\Omega t) 
\end{eqnarray}
for $i=1 \ldots N$. It is possible to understand in this case the origin of the resonance with the system size $N$, by deriving a closed equation for the collective (macroscopic) variable $\displaystyle X(t)=\frac{1}{N}\sum_{i=1}^N x_i $ as:
\begin{equation}
\label{eqmf}
\dot{X} = F(X) + \sqrt{\frac{D}{N}} \xi(t) + A \cos (\Omega t),
\end{equation}
where $\xi(t)$ is a zero mean Gaussian white noise with correlation function \makebox {$\langle \xi(t)\xi(t')\rangle =\delta(t-t')$}. The rescaling by $N^{-1/2}$ of the noise intensity has a simple origin in the central limit theorem. The function $F(X)$ can be computed by using some approximations based on the strong coupling limit, and it can be shown that it still exhibits bistable behavior~\cite{PZC02,TTMG}.

Eq. \ref{eqmf} shows that the effective noise intensity, $D/N$, can be controlled both by varying the noise intensity $D$ or the system size $N$. Hence, the optimal value of the effective noise intensity needed to observe stochastic resonance in the collective response $X(t)$ can be achieved by changing the system size $N$. It is then conceivable the following situation: let us start with a single system ($N=1$) subject to an external perturbation and noise, such that the noise intensity is too large in order to observe any synchrony with the weak external forcing, and the jumps between the two stable states occur randomly. If we now couple together an increasing number $N$ of these units, the effective noise for the global system will decrease as $N^{-1/2}$ and the global response to the external signal will be initially improved. Eventually, for too large $N$, the effective noise intensity will be very small and the system will be unable to follow globally the forcing. The possibility of having stochastic resonance for an optimal system size opens a wide range of applications in those cases in which it is not possible to tune the intensity of the noise at will, but it might be possible to change the number of coupled elements or the effective connections between them in order to obtain the best response.

In this paper we present an example of system size stochastic resonance in the field of the dynamics of social systems. Our objective is twofold. First, we want to show that the mechanism for system size stochastic resonance is generic and can appear in systems which are very far away from the original ones. Second, we emphasize the fact that the role of system size in social systems is a very important one and, for instance, there are examples which show that these systems may display phase transitions that {\sl disappear} in the thermodynamic limit, instead of the other way around which is the usual effect~\cite{TTAWS04}. 

The rest of the paper is organized as follows: in the next section we explain in some detail the model for opinion formation that we have considered in this work, while the results are presented in the final section \ref{section3}, and a short discussion in section \ref{section4}.

\section{Model studied}
\label{section2}

We have considered the model of opinion formation developed by Kuperman and Zanette~\cite{KZ02} based on similar models by Weidlich~\cite{W91}. In this model, the opinion is considered to be a binary variable, and we consider a set of $i=1,\dots,N$ individuals, each one having an opinion $\mu_i(t)=\pm 1$ at time $t$. 

The opinion of an individual is not fixed and it can change due to three effects: (i) the interaction with the rest of the individuals, modeled by a simple majority rule; (ii) the influence of fashion, modeled as the effect of some external time varying agent (such as advertising) and (iii) random changes. The model first establishes the connections between the individuals by enumerating the set $n(i)$ of neighbors of individual $i$.

In order to better mimic the social relations between the individuals, we assume that they live in the sites of a particular type of small-world network~\cite{WS98}. This is constructed as follows: we consider that the $i=1,\dots,N$ individuals are regularly spaced in a linear chain such that site $i$ is initially linked to the $2k$ nearest sites (we assume periodic boundary conditions). Each individual is then visited sequentially and with probability $p$ one of the links to its set of $k$ right near neighbors is randomly replaced by a link with a randomly chosen site. Double and multiple connections are forbidden, and realizations where the network becomes disconnected are discarded. In this way, the new set of neighbors $n(i)$ of site $i$, while still keeping an average size of $2k$, includes links to very far away sites. The {\sl re-wiring parameter} $p$ and the {\sl connectivity parameter} $k$ characterize the small-world network. 

The three effects mentioned above in the evolution of the opinion are precisely implemented as follows: assign at time $t=0$ random values $\mu_i=\pm 1$ to each individual; then at a given time $t$ the next three steps are applied consecutively:

\noindent{\bf (i)} Select randomly one individual $i$ and let it adopt the majority opinion favored by the set $n(i)$ of its neighbors, i.e. $\mu_i(t)={\rm sign} \left[\sum_{j\in n(i)}\mu_j(t)\right]$; if $\sum_{j\in n(i)}\mu_j(t)=0$ then $\mu_i(t)$ remains unchanged,\\
{\bf (ii)} with probability $A|\sin(\Omega t)|$, set $\mu_i(t)={\rm sign}\left[\sin(\Omega t)\right]$,\\
{\bf (iii)} with probability $\eta$, let $\mu_i$ adopt randomly a new value $\pm 1$, independently of its present value. 

After these three steps have been performed, time increases by a fixed amount $t\to t+1/N$. This is chosen such that after one unit of time every individual has been updated once on the average.

\begin{figure}
\begin{center}
\includegraphics[width=5cm,angle=-90]{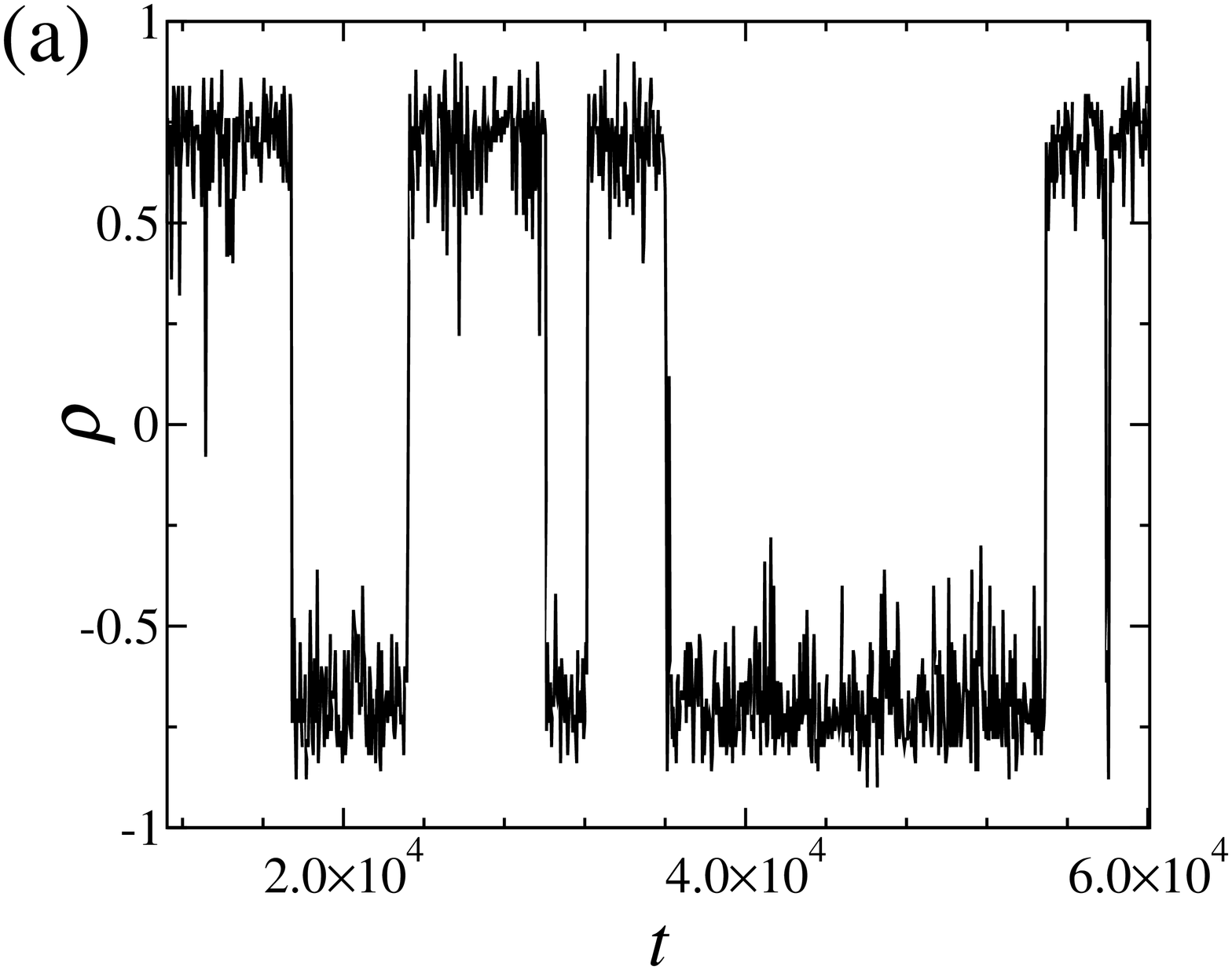}
\includegraphics[width=5cm,angle=-90]{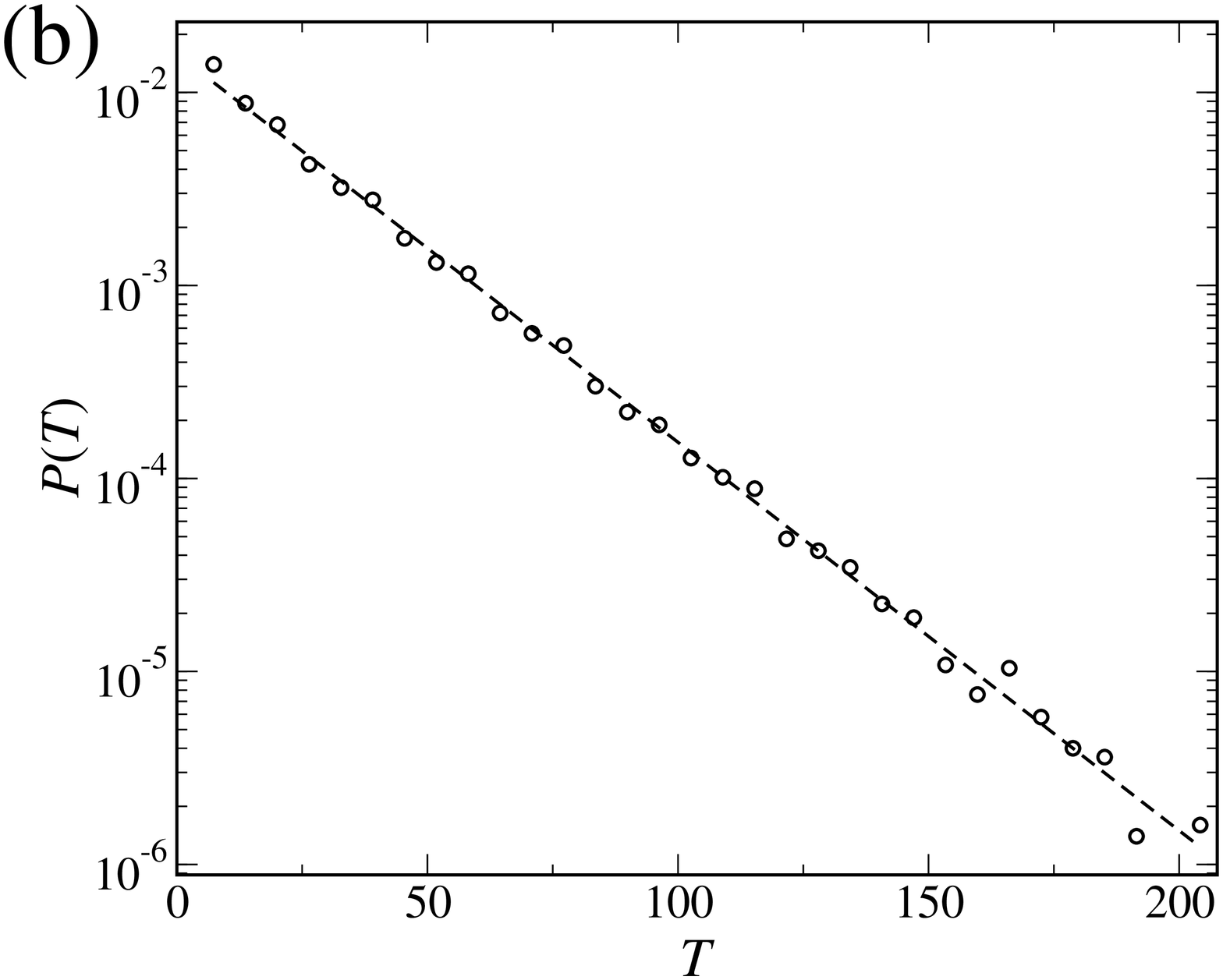}
\end{center}
\caption{\label{fig:dyn}(a) Time evolution of the average opinion $\rho(t)$ in the case $A=0$.\newline\hspace{1.0cm}(b) Distribution of the residence time $T$ on the stable states. The parameters are $k=3$, $p=0.3$, $N=100$ and $\eta=0.25$.}
\end{figure}

\begin{figure}
\begin{center}
\includegraphics[width=5cm,angle=-90]{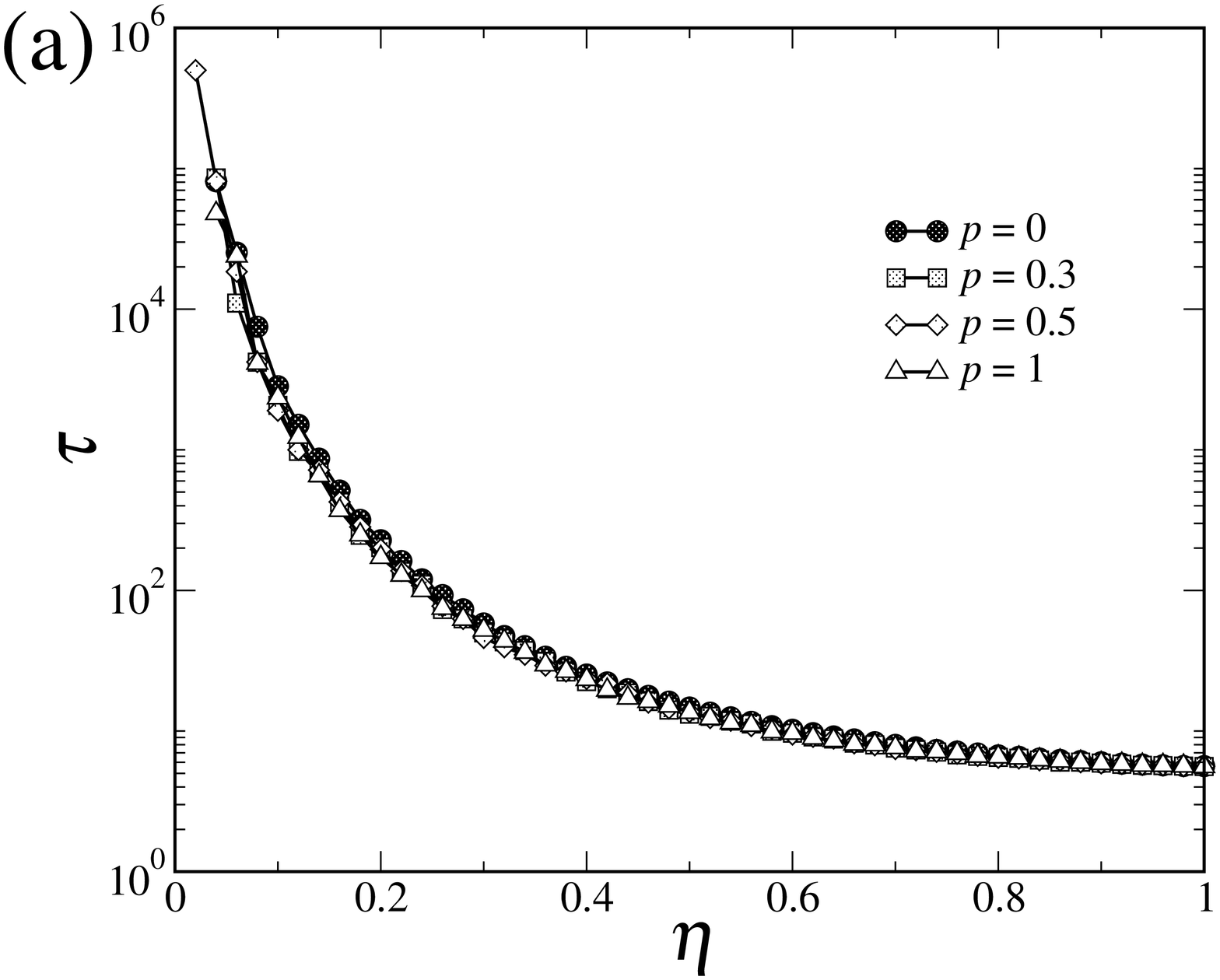}
\includegraphics[width=5cm,angle=-90]{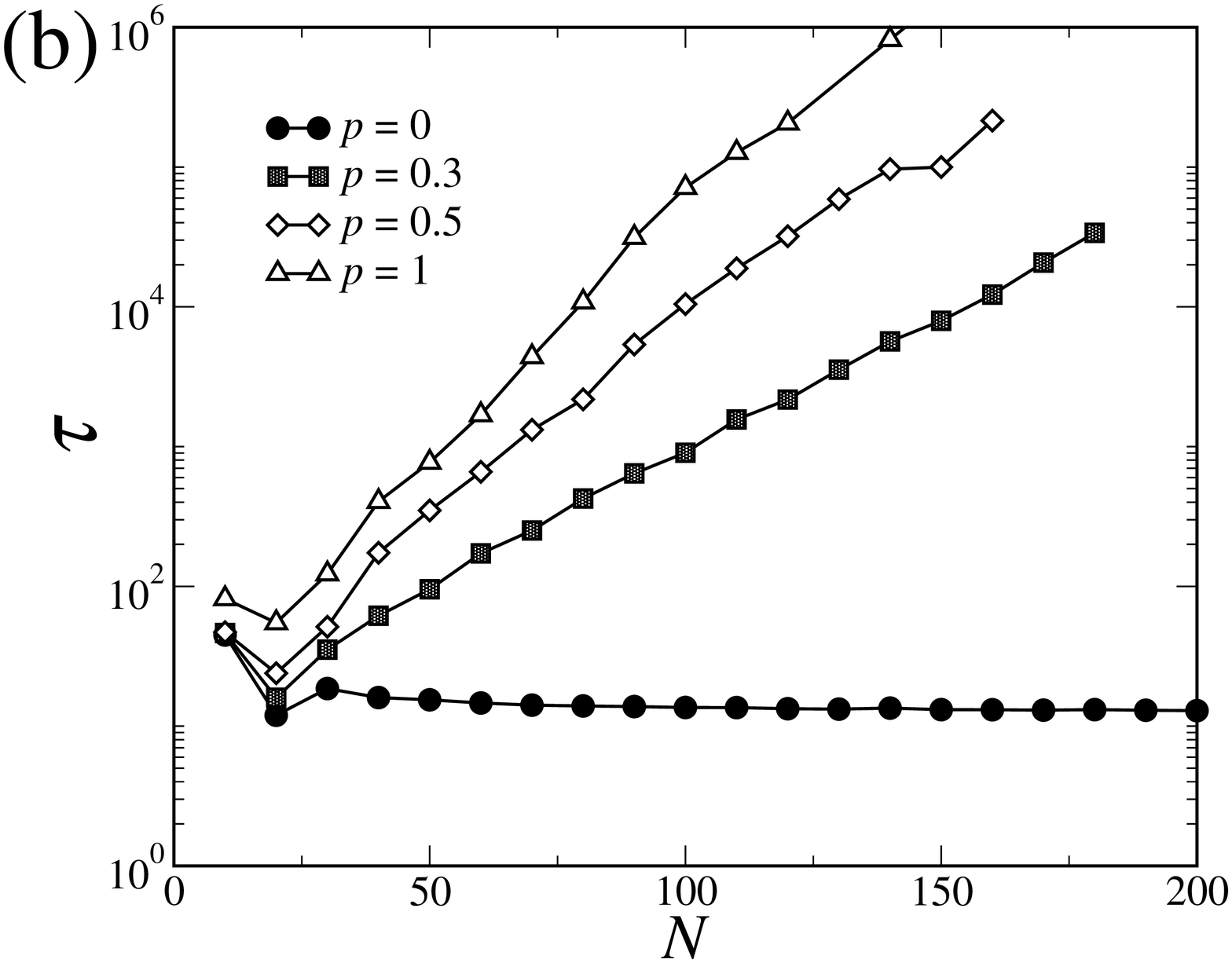}
\end{center}
\caption{\label{fig:tau-eta} (a) Mean residence time $\tau$ as a function of the flip rate $\eta$ in the case $N=100$.\newline(b) Mean residence time $\tau$ as a function of system size $N$ for $\eta=0.25$ and different values of the re-wiring parameter $p$. In both cases, other parameters are $A=0$, $k=3$.}
\end{figure}

The parameter $A$ ($0 \leq A \leq 1$) measures the strength of the fashion and $\Omega$ its frequency. The last step (iii) introduces noise in the evolution. In order to define a noise intensity $D$ related to the flip rate $\eta$, we consider the model without the effect of fashion (step ii). This is equivalent to setting $A=0$.  In figure \ref{fig:dyn} we plot the time dependence of the average opinion $\rho(t)=\frac{1}{N}\sum_i\mu_i(t)$. This figure clearly shows that the system behaves as bistable, jumping randomly around the two bistable states which are close to $\rho=\pm 1$. These random jumps are induced by the noise introduced in step (iii) of the evolution and occur more frequently for large flip rate $\eta$. This picture of a bistable system whose jumps between the stable states are induced by noise is consistent with the fact that, as shown in figure \ref{fig:dyn}(b), the residence time in each of the stable states follows the exponential Kramer's law:
\begin{equation}
p(T) = \tau \hbox{e}^{-T/\tau}
\end{equation}
being $\tau$ the mean residence time~\cite{kramers}. The dependence of $\tau$ in the flip rate $\eta$ and the system size $N$ can be seen in figure \ref{fig:tau-eta}.

\begin{figure}
\begin{center}
\includegraphics[width=6cm,angle=-90]{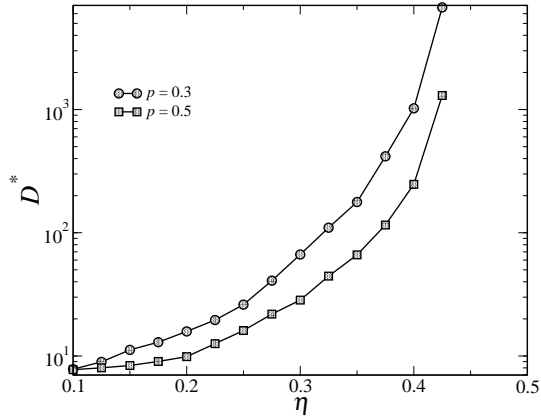}
\end{center}
\caption{\label{fig:deta}The rescaled noise $D^*=D/\Delta v$ as a function of flip rate $\eta$ for different values of the re-wiring parameter $p$ and $k=3$.}
\end{figure}

The next step is to define a noise intensity $D$ by using Kramer's formula, valid for small noise intensity,
\begin{equation}\label{eq:tauk}
\tau=\tau_0\hbox{e}^{\Delta V/D}
\end{equation}
being $\Delta V$ the height of the barrier between the two stable states. As shown in figure \ref{fig:tau-eta}(b), this barrier height increases with the number of individuals $\Delta V=N\Delta v$ with $\Delta v=\mathcal{O}(N^0)$, as expected. Thus, a simple fitting procedure allows us to obtain the rescaled noise intensity $D^*=D/\Delta v$. This is plotted in figure \ref{fig:deta} as a function of the flip rate $\eta$. 

\begin{figure}
\begin{center}
\includegraphics[width=5cm,angle=-90]{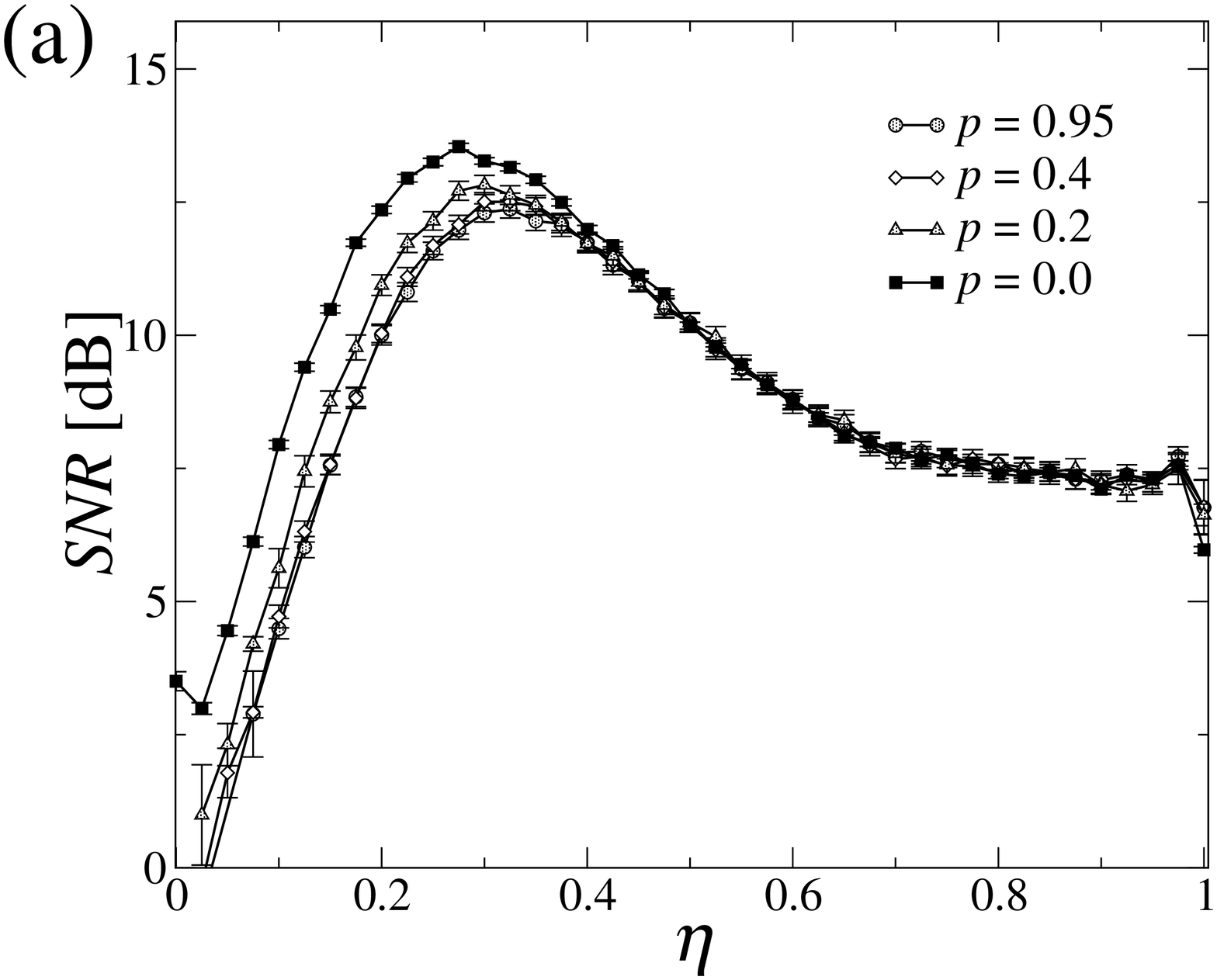}
\includegraphics[width=5cm,angle=-90]{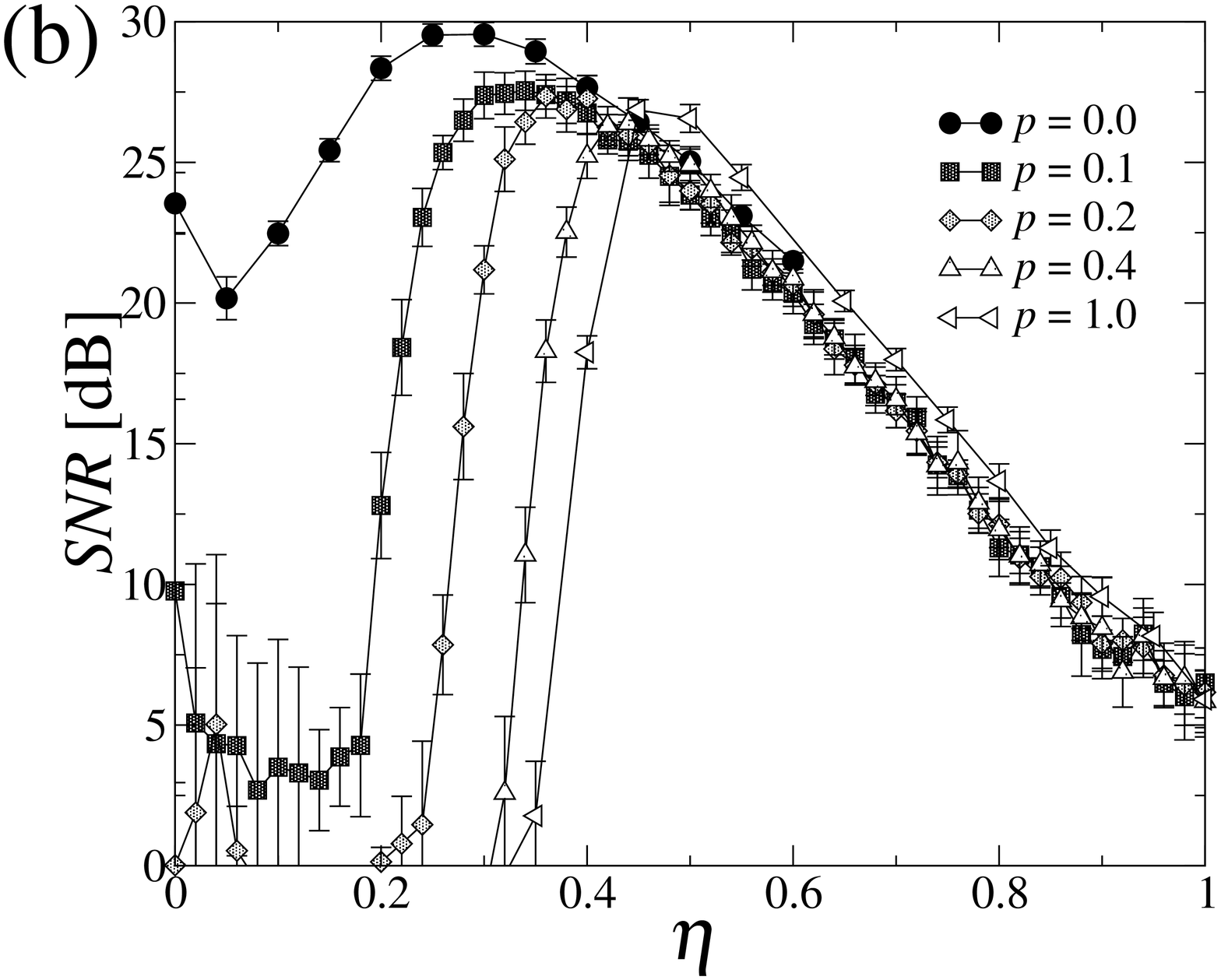}
\end{center}
\caption{\label{snr-d}The signal to noise ratio as a function of the flip rate $\eta$ for two different values of the system size: (a) $N=100$, and (b) $N=1024$. The presence of the maximum in each curve indicates the standard stochastic resonance phenomenon. The parameters are $k=3$, $p=0.3$, $\Omega=2\pi/128$ and $A=0.03$.}
\end{figure}

\section{System size stochastic resonance} 
\label{section3}

In the previous section we have shown that the opinion formation model considered is consistent with the picture of a bistable system with jumps between the two opinion states induced by the noise. In this section we turn our attention to the effect that the fashion, modeled as a periodic external signal, $A > 0$, has on the system. In particular, we ask ourselves the question of on which conditions the average opinion follows the fashion. Since the necessary ingredients are present in this model, it should not come to a surprise that this model displays stochastic resonance, as first shown in~\cite{KZ02}. A similar result was also found in~\cite{B97} for the original Weidlich model. The evidence is given in figures \ref{snr-d} which show that, for fixed values of $N$, the correlation between the majority opinion and the fashion is maximum for the proper value of the flip rate $\eta$. In these figures we plot the signal to noise ratio as a function of the flip rate $\eta$. In order to get a cleaner result, and as in other applications of stochastic resonance~\cite{GHJM98}, we have first filtered the original signal into a binary valued time series $s(t)={\rm sign}\left[\rho(t)\right]$. We then look at the power spectrum $S(\omega)$ of the time series of $s(t)$ and compute the signal to noise ratio in the standard form as the area above the background of $S(\omega)$ at the external frequency value $\omega=\Omega$.

\begin{figure}
\begin{center}
\includegraphics[width=11cm,angle=-90]{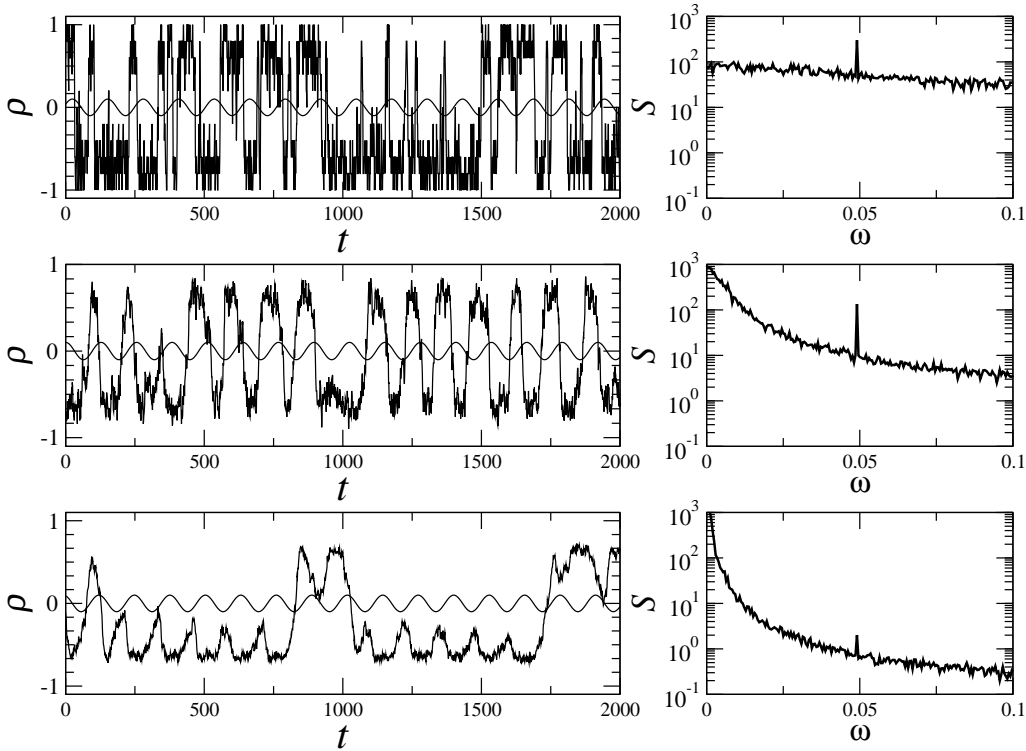}
\end{center}
\caption{\label{psddyn} Time series of the average opinion $\rho(t)$ for three different values of the system size: $N=10$ (upper figure), $N=100$ (middle) and $N=1000$ (lower). In each series we also plot a line proportional to the periodic function $\sin(\Omega t)$ modeling the influence of the fashion. Notice that the average opinion follows more closely the fashion for the intermediate value of $N$. For each of the time series, we also plot the power spectrum $S(\omega)$. Notice the peak at $\omega=\Omega$, the frequency of the fashion. The parameters of the system are $k=3$, $p=0.3$, $\Omega=2\pi/128$, $A=0.03$, $\eta=0.31$.}
\end{figure}

According to the general discussion in the introduction, we expect that the system will display as well stochastic resonance as a function of the number $N$ of individuals. This expectation is evidently fulfilled if one looks at the series of figures \ref{psddyn}. For small value of $N$ (upper figure), the average opinion $\rho(t)$ behaves rather erratically and independent of the periodic variation of the fashion. For a very large value of $N$ (lower figure), the average time between jumps is very large and, again, basically independent of the periodic variation of the fashion. It is only for an intermediate value of $N$ (middle figure) that the jumps between the two opinion states are correlated with the fashion. This result is also observed in the set of figures \ref{psddyn} which plot the power spectrum $S(\omega)$ coming from the corresponding time series. It is apparent in these figures that the signal to noise ratio first increases and then decreases when the number $N$ grows.

\begin{figure}
\begin{center}
\includegraphics[width=7cm,angle=-90]{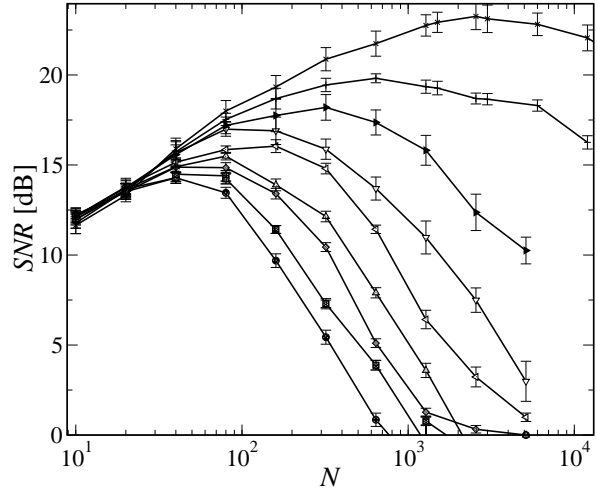}
\end{center}
\caption{\label{snr-n} The signal to noise ratio versus the system size $N$ for different values of flip-rate: $\eta=0.29, 0.30, 0.31, 0.32, 0.33, 0.34, 0.35, 0.36, 0.38$ from bottom to top. The existence of the maximum $N^*$ shows the system size stochastic resonance effect. Same parameters as is figure \ref{psddyn}.
}
\end{figure}

\begin{figure}
\begin{center}
\includegraphics[width=7cm,angle=-90]{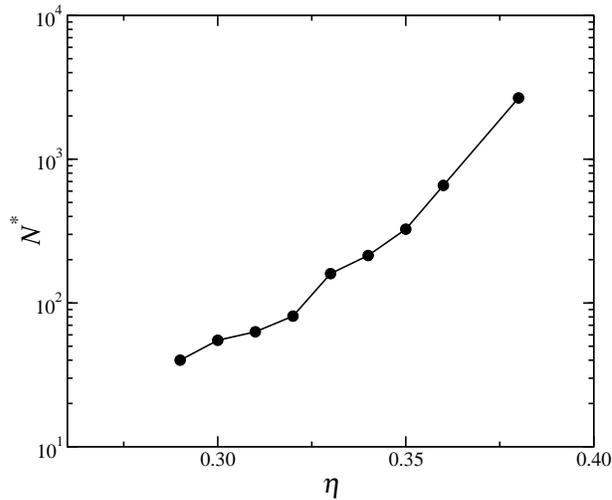}
\end{center}
\caption{\label{nstar-flip}The optimum value $N^*$ as a function of the flip rate $\eta$. Same parameters as is figure \ref{psddyn}.
}
\end{figure}

This main result is more clearly shown in figure \ref{snr-n} where we plot the signal to noise ratio as a function of the system size $N$ for different values of the flip rate. In each of the cases, it can be seen that there is an optimal value $N^*$ for which the signal to noise ratio takes it maximum value, indicating a maximum correlation between the average opinion and the fashion. The value of $N^*$ is plotted in figure \ref{nstar-flip} as a function of the flip rate $\eta$. 

\section{Conclusion}
\label{section4}
In conclusion, we have considered a model for opinion formation. The model incorporates three basic ingredients for the evolution of the opinion held by an individual: imitation, fashion and randomness. We have shown that in the absence of fashion, the model behaves as a bistable system with random jumps between the two stable states with a distribution of times following Kramer's law. We have used this image to compute the noise intensity as a function of the flip rate. Finally we have shown the existence of system size stochastic resonance, by which there is an optimal value for the number of individuals $N$ for which the average opinion follows better the fashion. This result indicates that the response of a social system to an external forcing agent depends in a non trivial manner of the  number of constituents, a feature already observed in other different models for social behavior.

{\bf Acknowledgments.} We thank M. Kuperman for fruitful discussions. We acknowledge financial support from the Ministerio de Ciencia y Tecnolog{\'\i}a (Spain) and FEDER through projects FIS2004-05073-C04-03 and FIS2004-00953.

\end{document}